\documentclass[showpacs,showkeys,preprintnumbers, amsmath,amssymb,preprint]{revtex4-1}
\usepackage{graphicx}
\usepackage{epsfig}
\usepackage{bm}

\usepackage{fontenc}
\usepackage{graphicx}
\usepackage[dvips]{hyperref}
\usepackage{bm}
\usepackage{amsmath}

\begin{document}

\title{Three-Dimensional Hairy Black Holes in Teleparallel Gravity}
\author{P. A. Gonz\'{a}lez}
\email{pablo.gonzalez@udp.cl}
\affiliation{Facultad de Ingenier\'{\i}a, Universidad Diego Portales, Avenida Ej\'{e}%
rcito Libertador 441, Casilla 298-V, Santiago, Chile.}
\author{Joel Saavedra}
\email{\,\,\,\,joel.saavedra@ucv.cl} \affiliation{Instituto de
F\'{i}sica, Pontificia Universidad Cat\'olica de Valpara\'{i}so,
Casilla 4950, Valpara\'{i}so, Chile.}
\author{Yerko V\'{a}squez.}
\email{yvasquez@userena.cl}
\affiliation{Departamento de F\'{\i}sica, Facultad de Ciencias, Universidad de La Serena,\\ 
Avenida Cisternas 1200, La Serena, Chile.}
\date{\today }

\begin{abstract}

We consider three-dimensional gravity based on torsion. Specifically, we consider an extension of the so-called Teleparallel Equivalent of General Relativity in the presence of a scalar field with a self-interacting potential, where the scalar field is non-minimally coupled with the torsion scalar. Then, we find asymptotically AdS hairy black hole solutions, which are characterized by a scalar field with a power-law behavior, being regular outside the event horizon and null at spatial infinity and by a self-interacting potential, which tends to an effective cosmological constant at spatial infinity.

\end{abstract}

\maketitle


\section{Introduction}

Although standard four-dimensional (4D) General Relativity (GR) is
believed to be the correct description of gravity at the classical level,
its quantization faces many well-known problems. Therefore,
three-dimensional (3D) gravity has gained much interest, since classically
it is much simpler and thus one can investigate more efficiently its
quantization. Amongst others, in 3D gravity one obtains the
Banados-Teitelboim-Zanelli (BTZ) black hole~\cite{BTZ}, which is a
solution to the Einstein equations with a negative cosmological constant.
This black-hole solution presents interesting properties at both classical
and quantum levels, and it shares several features of the Kerr black hole
of 4D GR~\cite{Carlip}. 

Furthermore, remarkable attention was addressed recently to topologically
massive gravity, which is a generalization of 3D GR that amounts to
augment the Einstein-Hilbert action by adding a Chern-Simons gravitational
term, and thus the propagating degree of freedom is a massive
graviton, which amongst others also admits BTZ black-hole as exact
solutions~\cite{deser}. The renewed interest on  topologically
massive gravity relies on the possibility of
constructing a chiral theory of gravity at a special point of the
parameter-space, as it was suggested in~\cite{Li:2008dq}. This idea has
been extensively analyzed in the last years~\cite{Strominger:2008dp}, leading to a fruitful
discussion that ultimately led to a significantly better understanding of
the model~\cite{Maloney:2009ck}. Moreover, another
3D massive gravity theory known as new massive gravity \cite{Bergshoeff:2009hq, Bergshoeff:2009tb} (where the action is given by the
Einstein-Hilbert term plus a specific
combination of square-curvature terms which gives rise to
field equations with a second order trace) have attracted considerable attention, this theory also admits interesting solutions, see for instance \cite{Clement:2009gq, Clement:2009ka, Oliva:2009ip}. Furthermore, 3D gravity with torsion has been extensively studied in~\cite{3dgravitywithtorsion} and references therein.


On the other hand, hairy black holes are interesting solutions of Einstein's Theory
of Gravity and also of certain types of Modified Gravity Theories. The first attempts to couple a scalar field to gravity was done in
an asymptotically flat spacetime. Then, hairy black hole solutions
were found \cite{BBMB} but these
solutions were not examples of hairy black hole configurations
violating the no-hair theorems because they were not physically
acceptable as the scalar field was divergent on the horizon and
stability analysis showed that they were unstable
\cite{bronnikov}. To remedy this a regularization procedure has to
be used to make the scalar field finite on the horizon. Hairy black hole solutions have been extensively studied over the years mainly in connection
 with the no-hair theorems. The recent developments in string theory and
specially the application of the AdS/CFT principle to condense
matter phenomena like superconductivity (for a review see
\cite{Hartnoll:2009sz}), triggered the interest of further  study
of the behavior of matter fields outside the black hole horizon
\cite{Gubser:2005ih,Gubser:2008px}. There are also very
interesting recent developments in Observational Astronomy. High
precision astronomical observations of the supermassive black
holes may pave the way to  experimentally test the no-hair
conjecture
 \cite{Sadeghian:2011ub}. Also, there are numerical investigations
 of single and binary black holes in the presence of scalar fields
\cite{Berti:2013gfa}. The aforementioned is a small part on the relevance that  has taken the study of hairy black holes currently in the field of physics,  for more details see for instance \cite{Gonzalez:2013aca, Gonzalez:2014tga} and references therein. Also, we refer the reader to references \cite{Martinez:1996gn, Henneaux:2002wm, Zhao:2013isa, Xu:2014uka, Cardenas:2014kaa} and references therein, where black holes solutions in three space-time dimensions with a scalar field (minimally and/or confomally) coupled to gravity have been investigated.



In the present work we are interested in investigating the existence of 3D hairy black holes solutions for theories based on torsion. In particular, the so-called ``teleparallel
equivalent of General Relativity" (TEGR) \cite{ein28,Hayashi79} is an
equivalent formulation of gravity but instead of using the curvature
defined via the Levi-Civita connection, it uses the Weitzenb{\"o}ck
connection that has no curvature but only torsion. So, we consider a scalar field non-minimally coupled with the torsion scalar, with a self-interacting potential in TEGR, and we find three-dimensional asymptotically AdS, hairy black holes. It is worth  mentioning, that this kind of theory (known as scalar-torsion theory), has been studied in the cosmological context, where the dark energy sector is attributed to the scalar field. It was shown that the minimal case is equivalent to standard quintessence. However,  the nonminimal case has a richer structure, exhibiting quintessence-like or phantom-like behavior, or experiencing the phantom-divide crossing \cite{Geng:2011aj, Geng:2011ka, Gu:2012ww}, see also \cite{Horvat:2014xwa} for aplications of this theory (with a complex scalar field) to boson stars.

It is also worth to mention that a natural extension of TEGR is the so called $f(T)$ gravity, which is represented by a function of the scalar torsion $T$ as Lagrangian density \cite{Ferraro:2006jd, Ferraro:2008ey, Bengochea:2008gz,Linder:2010py}.
The $f(T)$ theories picks up preferred referential frames which constitute the autoparallel curves of the given manifold. 
A genuine advantage of $f(T)$ gravity  compared with other deformed gravitational schemes is that the differential equations for the vielbein components are second order differential equations. However, the effects of the additional degrees of freedom that certainly exist in $f(T)$ theories is a consequence of breaking the local Lorentz invariance that  these theories exhibit. Despite this, it was found that on the flat FRW background with a scalar field,  up to second order linear perturbations  does not reveal any extra degree of freedom at all \cite{Izumi:2012qj}. As such, it is fair to say  that the nature of these additional degrees of freedom remains unknown. Remarkably, it is possible to modify $f(T)$ theory in order to make it manifestly Lorentz invariant. However, it will generically have different dynamics and will reduce to $f(T)$ gravity in some local Lorentz frames \cite{Li:2010cg, Weinberg, Arcos:2010gi}.
Clearly, in extending this
geometry sector, one of the goals is to solve the puzzle of dark energy and
dark matter without asking for new material ingredients that have
not yet been detected  by  experiments \cite{Capozziello:2007ec,Ghosh:2012pg}. For instance, a Born-Infeld $f(T)$ gravity Lagrangian was used to address the physically inadmissible divergencies occurring in the standard cosmological Big Bang model, rendering the spacetime geodesically complete and powering an inflationary stage without the introduction of an inflaton field \cite{Ferraro:2008ey}. Also, it is believed that $f(T)$ gravity could be a reliable approach to address the shortcomings of general relativity at high energy scales \cite{Capozziello:2011et}.  Furthermore, both inflation and the dark energy dominated stage can be realized in Kaluza-Klein and Randall-Sundrum models, respectively \cite{Bamba:2013fta}.
In this way, $f(T)$ gravity has gained attention and
has been proven to exhibit interesting cosmological implications. On the other hand, the search for black hole solutions in $f(T)$ gravity is not a trivial problem, and there are only few exact solutions, see for instance \cite{G1, solutions,Rodrigues:2013ifa}. Remarkably, it is possible to construct other generalizations, as Teleparallel Equivalent of Gauss-Bonnet Gravity \cite{Kofinas:2014owa, Kofinas:2014daa}, Kaluza-Klein theory for teleparallel gravity \cite{Geng:2014nfa} and scalar-torsion gravity theories \cite{Geng:2011aj, Kofinas:2015hla}.

The paper is organized as follows. In Section II we give a brief review of three-dimensional Teleparallel Gravity. Then, in Section III we find asymptotically AdS black holes with scalar hair, and we conclude in Section IV with final remarks.

\section{3D Teleparallel  Gravity}
\label{Tel3D}

In 1928, Einstein proposed the idea of teleparallelism to unify gravity and electromagnetism into a unified field theory; this corresponds to an equivalent formulation of General Relativity (GR), nowadays known as Teleparallel Equivalent to General Relativity (TEGR) \cite{ein28, Hayashi79}, where the Weitzenb\"{o}ck connection is used  to define the covariant derivative (instead of the Levi-Civita connection which is used to define the covariant derivative in the context of GR). The first 
 investigations on teleparallel 3D gravity were
performed by Kawai almost twenty years ago \cite{Kawai1,Kawai2,Kawai3}. The Weitzenb\"{o}ck connection mentioned above has not null torsion. However, it is curvatureless, which implies that this formulation of gravity exhibits only torsion. The Lagrangian density $T$ is constructed from the torsion tensor.
To clarify, the torsion scalar $T$ is the result of a very specific quadratic combination of irreducible representations of the torsion tensor under the Lorentz group $SO(1,3)$ \cite{Hehl:1994ue}. In this way,  the torsion tensor in TEGR  
includes all the information concerning  the
gravitational field.
The theory is called ``Teleparallel
Equivalent to General Relativity'' since the field equations are exactly the same as those of GR for every geometry choice.

The Lagrangian of teleparallel 3D gravity corresponds to
the more general quadratic Lagrangian for torsion, under the
assumption of zero spin-connection. So, the action can be written as \cite{Muench:1998ay,Itin:1999wi}
\begin{equation}  \label{action2}
S=\frac{1}{2 \kappa}\int \left( \rho_{0} \mathcal{L}_{0}+ \rho_{1}
\mathcal{L%
}_{1}+ \rho_{2} \mathcal{L}_{2}+\rho_{3} \mathcal{L}_{3}+ \rho_{4}
\mathcal{L%
}_{4}\right)~,
\end{equation}%
where $\kappa$ is the three-dimensional gravitational constant,
$\rho_i$ are parameters, and
\begin{equation}
\mathcal{L}_{0}= \frac{1}{4}e^{a} \wedge \star e_a~,\quad \mathcal{L}_{1}=de^{a}
\wedge \star de_{a}~,\quad \mathcal{L}_{2}= (de_{a} \wedge \star e^a)
\wedge \star (de_b \wedge e^b)~,\nonumber
\end{equation}
\begin{equation}
\mathcal{L}_{3}=(de^{a} \wedge e^{b}) \wedge \star (de_{a} \wedge
e_{b})~,\quad \mathcal{L}_{4}= (de_{a} \wedge \star e^b) \wedge
\star (de_b \wedge e^a)~,
\end{equation}
where $e^a$ denotes the vielbein, $d$ is the exterior derivative, $\star $ denotes the Hodge dual operator and $\wedge$ the wedge
product. The coupling constant $\rho_{0}=-\frac{8}{3} \Lambda$ represents
the cosmological constant term.  Moreover, since $\mathcal{L}_{3}$ can be
written completely in terms of $\mathcal{L}%
_{1}$, in the following we set $\rho_{3}=0$ \cite{Muench:1998ay}.  Action (\ref{action2}) can be written in a more convenient form as
\begin{equation}
\label{actiontele0}
S=\frac{1}{2\kappa} \int \left (T -2\Lambda   \right )\star 1~,
\end{equation}
where $\star1=e^{0} \wedge e^{1} \wedge e^{2}$, and the torsion
scalar $T$ is given by
\begin{equation}  \label{scalartorsion}
T= \star \left[\rho_{1}(de^{a} \wedge \star de_{a})+\rho_{2}(de_{a} \wedge
e^a) \wedge \star (de_b \wedge e^b)+\rho_{4}(de_{a} \wedge
e^b) \wedge \star (de_b \wedge e^a) \right]~.
\end{equation}
Expanding this expression in terms of its components, the torsion scalar yields
\begin{equation}
\label{scalartorsionrho}
T=\frac{1}{2} (\rho_{1}+\rho_{2}+\rho_{4})T^{abc}T_{abc}+%
\rho_{2}T^{abc}T_{bca}-\rho_{4}T_{a}^{ac}T^{b}_{bc}~,
\end{equation}
note that for TEGR $\rho_{1}=0$, $\rho_{2}=-\frac{1}{2}$ and $\rho_{4}=1$. 
A variation of  action (\ref{actiontele0}) with respect to the
vielbein provides the following field equations:
\begin{eqnarray}\label{fieldequations}
&&\delta \mathcal{L} =\delta e^{a}\wedge \left\{\left\{\rho
_{1}\left[2d\star de_{a}+i_{a}(de^{b}\wedge \star
de_{b})-2i_{a}(de^{b})\wedge
\star de_{b}\right]
\right.\right.\nonumber
\\ && \ \ \ \ \ \ \ \ \ \ \ \ \ \ \ \ \ \
+\rho _{2}\left\{-2e_{a}\wedge d\star (de^{b}\wedge
e_{b})
+2de_{a}\wedge \star (de^{b}\wedge e_{b})+i_{a}\left[de^{c}\wedge
e_{c}\wedge
\star (de^{b}\wedge e_{b})\right]\right.\nonumber\\
&&\ \ \ \ \ \ \ \ \ \ \ \ \ \ \ \ \ \ \ \ \ \, \ \ \ \ \ \left.
-2i_{a}(de^{b})\wedge e_{b}\wedge \star
(de^{c}\wedge e_{c})\right\}
 \nonumber\\
&& \ \ \ \ \ \ \ \ \ \ \ \ \ \ \ \ \ \ +\rho_{4}\left\{-2e_{b}\wedge
d\star (e_{a}\wedge de^{b})+2de_{b}\wedge \star
(e_{a}\wedge de^{b})+i_{a}\left[e_{c}\wedge de^{b}\wedge \star
(de^{c}\wedge
e_{b})\right]\right.\nonumber \\
&&\ \ \ \ \ \ \ \ \ \ \ \ \ \ \ \ \ \ \ \ \ \, \ \ \ \ \ \left.\left.
 -2i_{a}(de^{b})\wedge e_{c}\wedge \star (de^{c}\wedge
e_{b})\right\}\right\}\nonumber\\
&&\ \left. \ \ \ \ \ \ \ \ \ \ \ \ \ \ \ \
-2\Lambda \star e_a  \right\}
=0~,
\end{eqnarray}
where $i_a$ is the interior product and for generality's sake we have kept the
general
coefficients $\rho_i$, and we have used $\epsilon^{012}=+1$. Through the following choice of the coefficients $\rho_{1}=0$, $\rho_{2}=-\frac{1}{2}$ and $\rho_{4}=1$ Teleparallel Gravity  coincides with the
usual curvature-formulation of General Relativity and therefore the following BTZ metric is solution of TEGR  
 \begin{equation}
  \label{metric}
ds^2=N^2dt^2-N^{-2}dr^2-r^2(d\varphi+N_{\varphi}dt)^2~,
\end{equation}
where the lapse $N$ and shift $N_{\varphi}$ functions are given by,
\begin{equation}
N^2= -8GM+\frac{r^2}{l^2}+\frac{16G^2J^2}{r^2}~,\quad
N_{\varphi}=-\frac{4GJ}{r^2},
\label{BTZ0}
\end{equation}
and the two constants of integration $M$ and $J$ are the usual conserved
charges associated with the asymptotic invariance under time displacements
(mass) and rotational invariance (angular momentum)
respectively,  given by flux integrals
through a large circle at spacelike infinity, and $\Lambda=-1/l^2$ is the
cosmological constant \cite{BTZ}.
Finally, note
that 
the torsion scalar can be  
calculated, leading to
the constant value
\begin{equation}
\label{Tteleresult}
T=-2\Lambda,
\end{equation}
which is the cosmological constant as the sole source of torsion.\\

\section{3D Teleparallel Hairy Black Holes}
\label{Tel3DH}
\subsection{The Model}

In this section we will extend the above discussion considering a scalar field $\phi$ non-minimally coupled with the torsion scalar with a self-interacting potential $V(\phi)$, and then we will find hairy black hole solutions. 
 So, the action can be written as 
\begin{equation}
  \label{accionHT}
S=\int \left( \frac{1}{2 \kappa} T \star 1 - \xi \phi^2 T \star 1 + \frac{1}{2} d\phi \wedge \star d\phi -V(\phi)\star 1\right)~,
\end{equation}
where $T$ is given by (\ref{scalartorsion}) and $\xi$ is the non-minimal coupling parameter. 
Thus, the variation with respect to the vielbein leads to
 the following field equations:
\begin{eqnarray}  \label{fieldeq}
\delta_{e^{a}} \mathcal{L} &=&\delta e^{a}\wedge
\left\{\left(\frac{1}{2\kappa}-\xi\phi^2\right)
\left\{\rho
_{1}\left[2d\star de_{a}+i_{a}(de^{b}\wedge \star
de_{b})-2i_{a}(de^{b})\wedge
\star de_{b}\right]\right.\right.\nonumber
\\
&&\ \ \ \ \ \ \ \ +\rho _{2}\left\{-2e_{a}\wedge d\star (de^{b}\wedge
e_{b})+2de_{a}\wedge \star
(de^{b}\wedge e_{b})+i_{a}\left[de^{c}\wedge e_{c}\wedge
\star (de^{b}\wedge e_{b})\right] \right.
\nonumber\\
&&\left. \ \ \ \ \ \ \ \  \ \ \ \ \ \ \ \,  -2i_{a}(de^{b})\wedge
e_{b}\wedge \star
(de^{c}\wedge e_{c})\right\}\nonumber\\
&&\ \ \ \ \ \ \ \ 
+\rho_{4}\left\{-2e_{b}\wedge d\star (e_{a}\wedge
de^{b})+2de_{b}\wedge \star
(e_{a}\wedge de^{b})\right.
\nonumber \\
&&\left.\left.\ \ \ \ \ \ \ \  \ \ \ \ \ \ \ \,
+i_{a}\left[e_{c}\wedge de^{b}\wedge \star (de^{c}\wedge
e_{b})\right] -2i_{a}(de^{b})\wedge e_{c}\wedge \star (de^{c}\wedge
e_{b})\right\}
\right\}\nonumber\\
&&\ \ \ \ \ \ \ \ -4\xi\left[\rho_1\phi d\phi \wedge \star de_a+\rho_2 \phi d\phi\wedge
e_a\wedge
\star(de_b\wedge e^b)+\rho_4\phi d\phi \wedge e_b\wedge \star(de^b \wedge
e_a)\right]\nonumber\\
&&\left. \ \ \ \ \ \ \ \ - V(\phi) i_a(\star 1)-\frac{1}{2}d\phi\wedge i_a(\star d\phi)-\frac{1}{2} i_a(d\phi)\wedge\star d\phi    \right\}=0~,
\label{fieldeq000}
\end{eqnarray}
and the variation with respect to the scalar field leads to the Klein-Gordon equation 
\begin{equation}
\label{fieldeq001}
\delta_{\phi} \mathcal{L}=\delta \phi \left(-2\xi\phi T \star 1-d\star d\phi - \frac{dV}{d\phi}\star 1 \right)=0~.
\end{equation}

\subsection{Circularly Symmetric Hairy Solutions}
\label{circsymmsol}
Let us now investigate hairy black hole solutions of the theory. In order to
analyze static solutions we consider
the metric form as
\begin{equation}\label{metric}
ds^{2}=A\left( r\right) ^{2}dt^{2}-\frac{1}{B\left( r\right) ^{2}}
dr^{2}-r^{2}d\varphi^{2}~,
\end{equation}
which arises from the triad 
diagonal ansatz 
\begin{equation}\label{diagonal}
e^{0}=A\left( r\right) dt~,\text{ \ }e^{1}=\frac{1}{B\left( r\right) }dr~,
\text{ \ }e^{2}=rd\varphi~.
\end{equation}
Then, inserting this vielbein in the field equations (\ref{fieldeq000}), (\ref{fieldeq001}) yields
\begin{equation}
-\frac{1}{r}{(\frac{1}{2\kappa}-\xi\phi(r)^2)\frac{dB^2}{dr}}+\frac{4}{r}\xi B(r)^2\phi(r)\frac{d\phi}{dr}-\frac{1}{2}B(r)^2(\frac{d\phi}{dr})^2-V(\phi)=0~,
\label{q1}
\end{equation}
\begin{equation}
\frac{B(r)^2}{rA(r)^2}{(\frac{1}{2\kappa}-\xi\phi(r)^2)\frac{dA^2}{dr}}-\frac{1}{2}B(r)^2(\frac{d\phi}{dr})^2+V(\phi)=0~, 
\label{q2}
\end{equation}
\begin{eqnarray}
\notag&& 2 \xi\phi(r) \frac{d\phi}{dr}\frac{B(r)^2}{A(r)^2}\frac{dA^2}{dr}-\frac{1}{2}B(r)^2(\frac{d\phi}{dr})^2\\
\notag&&+\frac{1}{2A(r)^4}(\frac{1}{2\kappa}-\xi\phi(r)^2)\left(-A(r)^2\frac{dA^2}{dr}\frac{dB^2}{dr}+B(r)^2(\frac{dA^2}{dr})^2-2A(r)^2 B(r)^2\frac{d^2A^2}{dr^2}\right)\\
&&-V(\phi)=0~,
\label{q3}
\end{eqnarray}
\begin{equation}
-\frac{2B(r)^2}{rA(r)^2}\xi\phi(r)\frac{dA^2}{dr}+\frac{1}{r}B(r)^2\frac{d\phi}{dr}+\frac{1}{2}\frac{dB(r)^2}{dr}\frac{d\phi}{dr}+\frac{B(r)^2}{2A(r)^2}\frac{dA(r)^2}{dr}\frac{d\phi}{dr}+B(r)^2\frac{d^2\phi}{dr^2}-\frac{dV}{d\phi}=0~.
\label{q4}
\end{equation}

It is worth mentioning that, in the case of a minimally coupled scalar field,  
 the above  simple, diagonal  relation 
 between the metric and the vielbeins (\ref{diagonal}) is
always allowed, due to in this case the theory is invariant under local Lorentz transformations of the vielbein. In contrast, in the extension of a non-minimally coupled scalar field with the torsion scalar, the theory is not local Lorentz invariant, therefore, one could have in general a more complicated relation connecting the vielbein with the metric, with the vielbeins being non-diagonal even for a diagonal
metric \cite{fTLorinv0}. However, for the three-dimensional solutions considered here, using a preferred diagonal frame is allowed, in the sense that this frame defines a global set of basis covering the whole tangent bundle, i.e., they parallelize the spacetime \cite{Fiorini:2013hva}, \cite{Ferraro:2011us}.

In the following, and in order to solve the above system of equations, we will consider two cases:  first, we analyze the case $A(r)=B(r)$, and then we analyze the more general case $A(r) \neq B(r)$.
\subsubsection{$A(r)=B(r)$}
\begin{figure}[h]
\begin{center}
\includegraphics[width=0.6\textwidth]{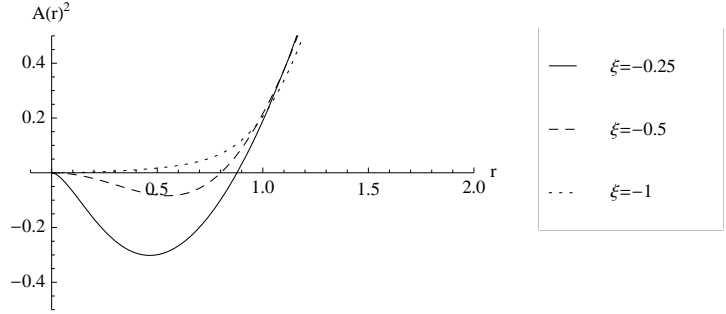}
\end{center}
\caption{The behavior of $A(r)^2$, for $H=-1$, $G=1$, $B=1$, $\kappa=1$, 
and $\xi=-0.25,-0.5,-1$.} \label{function}
\end{figure}

\begin{figure}[h]
\begin{center}
\includegraphics[width=0.6\textwidth]{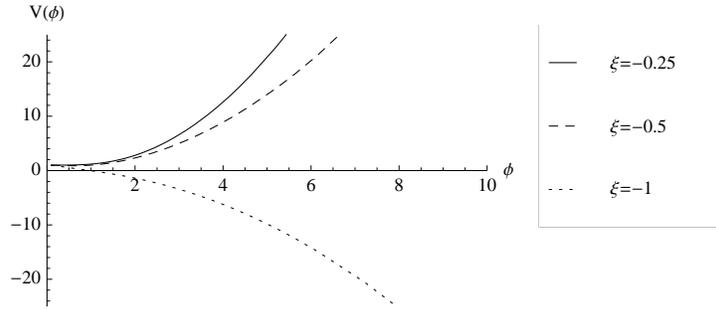}
\end{center}
\caption{The Potencial $V(\phi)$, for $H=-1$, $G=1$, $B=1$, $\kappa=1$, 
and $\xi=-0.25,-0.5,-1$.} \label{Pot1}
\end{figure}

\begin{figure}[h]
\begin{center}
\includegraphics[width=0.4\textwidth]{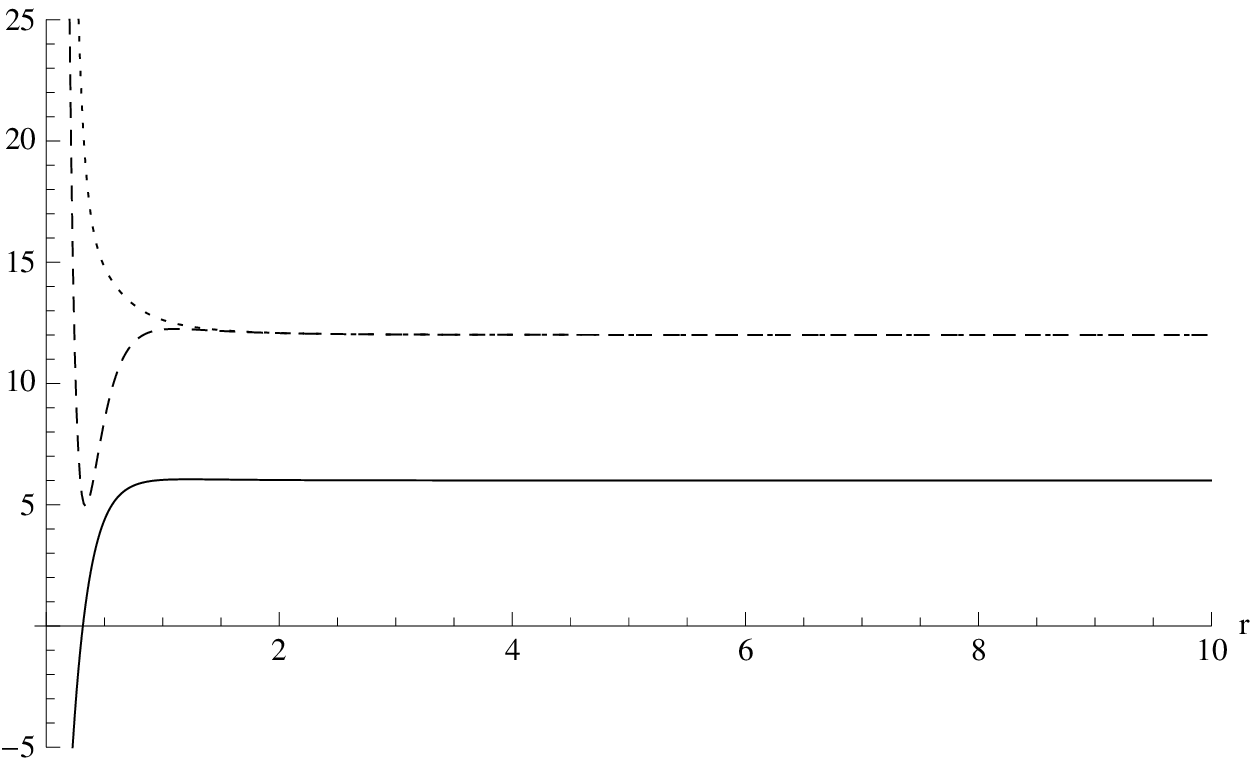}
\includegraphics[width=0.4\textwidth]{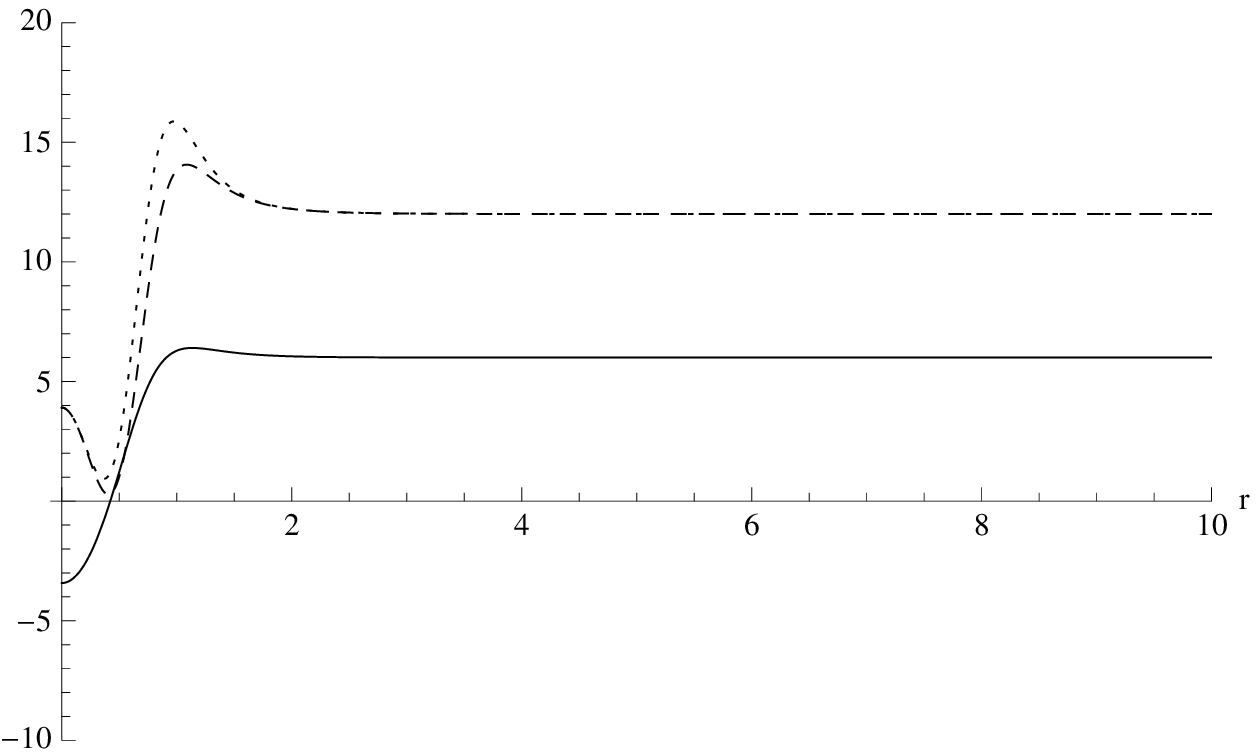}
\includegraphics[width=0.55\textwidth]{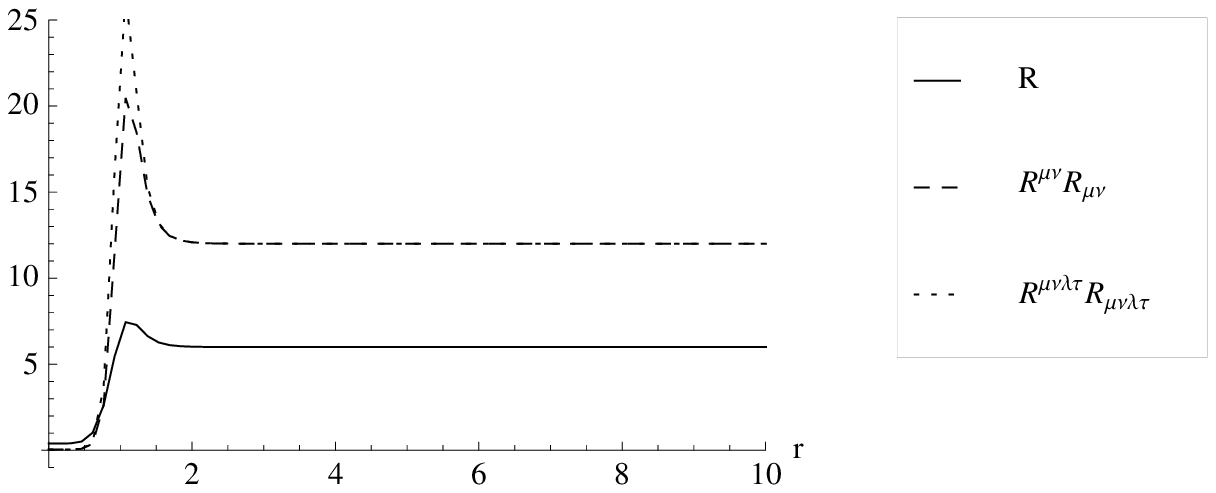}
\end{center}
\caption{The behavior of $R(r)$, $R^{\mu\nu}R_{\mu\nu}(r)$ and $R^{\mu\nu\lambda\tau}R_{\mu\nu\lambda\tau}(r)$ for $H=-1$, $G=1$, $B=1$, $\kappa=1$, and $\xi=-0.25$ (left figure), $\xi=-0.5$ (right figure), and $\xi=-1$ (bottom figure).} \label{figuraRR}
\end{figure}

\begin{figure}[h]
\begin{center}
\includegraphics[width=0.6\textwidth]{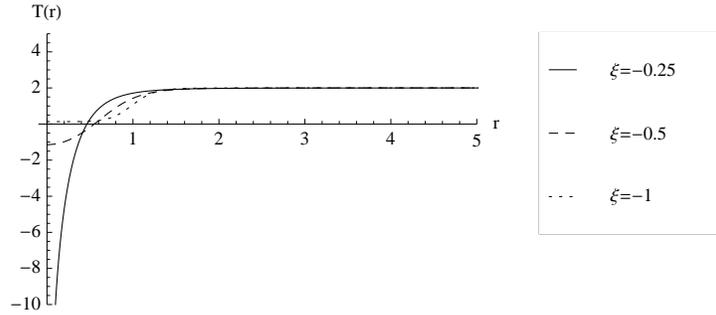}
\end{center}
\caption{The behavior of torsion scalar $T$ as function of $r$ for $H=-1$, $G=1$, $B=1$, $\kappa=1$, and $\xi=-0.25, -0.5, -1$.} \label{figuraR}
\end{figure}

In this case the field equations (\ref{q1})-(\ref{q4}) simplify to
\begin{equation}
-\frac{1}{r}{(\frac{1}{2\kappa}-\xi\phi(r)^2)\frac{dA^2}{dr}}+\frac{4}{r}\xi A(r)^2\phi(r)\frac{d\phi}{dr}-\frac{1}{2}A(r)^2(\frac{d\phi}{dr})^2-V(\phi)=0~, 
\label{fieldequation1}
\end{equation}
\begin{equation}
\frac{1}{r}{(\frac{1}{2\kappa}-\xi\phi(r)^2)\frac{dA^2}{dr}}-\frac{1}{2}A(r)^2(\frac{d\phi}{dr})^2+V(\phi)=0~, 
\label{fieldequation2}
\end{equation}
\begin{equation}
2\xi\phi(r) \frac{d\phi}{dr}\frac{dA^2}{dr}-\frac{1}{2}A(r)^2(\frac{d\phi}{dr})^2-(\frac{1}{2\kappa}-\xi\phi(r)^2)\frac{d^2A^2}{dr^2}-V(\phi)=0~,
\label{fieldequation3}
\end{equation}
\begin{equation}
-\frac{2}{r}\xi\phi(r)\frac{dA^2}{dr}+\frac{1}{r}A(r)^2\frac{d\phi}{dr}+\frac{dA(r)^2}{dr}\frac{d\phi}{dr}+A(r)^2\frac{d^2\phi}{dr^2}-\frac{dV}{d\phi}=0~.
\label{fieldequation4}
\end{equation}
Now, by adding equations (\ref{fieldequation1}) and (\ref{fieldequation2}) we obtain
\begin{equation}
A(r)^2\frac{d \phi}{dr}\left( \frac{4 \xi}{r} \phi-\frac{d \phi}{dr}\right)=0~.
\end{equation}
Therefore, the nontrivial solution for the scalar field is given by
\begin{equation}
\phi(r)=Br^{4\xi} ~,
\end{equation}
and by using this profile for the scalar field in the remaining equations, we obtain the solution
\begin{equation}
\label{A}
A(r)^2=Gr^{2}+H {}_{2}F_1(1,-\frac{1}{4\xi}, 1-\frac{1}{4\xi}, 2\kappa B^2\xi r^{8\xi})~,
\end{equation}
\begin{eqnarray}
V(\phi) & = & \frac{H}{\kappa}\left(\frac{\phi}{B}\right)^{-\frac{1}{2\xi}}+2G\left(-\frac{1}{2\kappa}+B^2\xi(1+4\xi)\left(\frac{\phi}{B}\right)^2\right) \\ \notag 
&& -2H\left(\frac{\phi}{B}\right)^{-\frac{1}{2\xi}}\left(\frac{1}{2\kappa}-B^2\xi(1+4\xi)\left(\frac{\phi}{B}\right)^2\right){}_{2}F_1(1,-\frac{1}{4\xi}, 1-\frac{1}{4\xi}, 2\kappa \xi \phi^2)~,
\end{eqnarray}
where $B$, $G$, and $H$ are integration constant and ${}_{2}F_{1}$ is the Gauss hypergeometric function. In the limits $\xi \rightarrow 0$ or $B \rightarrow 0$ the theory reduces to TEGR, therefore, we must hope our solution reduces to the BTZ black hole, this is indeed the case, as we show below. For those limits we obtain:
\begin{eqnarray}
\lim_{\xi \rightarrow 0} \,\,{}_{2}F_1(1,-\frac{1}{4\xi}, 1-\frac{1}{4\xi}, 2\kappa B^2\xi r^{8\xi})=1~,\\
\lim_{B \rightarrow 0} \,\,{}_{2}F_1(1,-\frac{1}{4\xi}, 1-\frac{1}{4\xi}, 2\kappa B^2\xi r^{8\xi})=1~,
\end{eqnarray}
therefore,
\begin{equation}
\lim_{\xi \rightarrow 0 \,\, or \,\, B \rightarrow 0} \,\, A(r)^2=Gr^{2}+H~,
\end{equation}
which is the non-rotating BTZ metric.
In order to see the asymptotic behavior of $A(r)^2$, we expand the hypergeometric function for large $r$ and $\xi<0$:
\begin{equation}
{}_{2}F_1(1,-\frac{1}{4\xi}, 1-\frac{1}{4\xi}, 2\kappa B^2\xi r^{8\xi})\approx 1-\frac{\kappa B^2 r^{8 \xi}}{2 \left( 1-\frac{1}{4 \xi}\right)}+...~.
\end{equation}
This expansion shows that the hairy black hole is asymptotically AdS.
On other hand, in the limit $\phi\rightarrow 0$ the potential goes to a constant (the effective cosmological constant) $V(\phi)\rightarrow -\frac{G}{\kappa}=\Lambda$. In Fig.~(\ref{function}) we plot the  behavior of the metric function
$A\left( r\right)^2 $ given by Eq. (\ref{A}) 
for a choice of parameters $H=-1$, $G=1$, $B=1$, $\kappa=1$, 
and $\xi=-0.25,-0.5,-1$. The metric function $A(r)^2$ changes
sign for low values of $r$, signalling the presence of a horizon,
while the scalar field is
regular everywhere outside the event horizon (for $\xi < 0$) and null at large
distances. In Fig.~(\ref{Pot1}) we show the behavior of the potential, and we observe that it tends asymptotically ($\phi \rightarrow 0$)  to a negative constant (the effective cosmological constant). We also plot the
behavior of the Ricci scalar $R(r)$, the principal quadratic invariant of the Ricci tensor $R^{\mu\nu}R_{\mu\nu}(r)$, and the Kretschmann scalar
$R^{\mu\nu\lambda\tau}R_{\mu\nu\lambda\tau}(r)$  in Fig.~(\ref{figuraRR}) by using the Levi-Civita connection, and we observe that there is not a
Riemann curvature singularity outside the horizon for  $\xi=-0.25,-0.5,-1$. Also, we observe a Riemann curvature singularity at $r=0$ for $\xi=-0.25$ and the torsion scalar is singular at $r=0$ for $\xi=-0.25$, see Fig.~(\ref{figuraR}).

\subsubsection{$A(r) \neq B(r)$}

Now, by considering the following ansatz for the scalar field
\begin{equation}
\phi(r)=Br^{\gamma} ~,
\end{equation}
we find the following solution to the field equations
\begin{equation}
A(r)^2=Gr^{2}+H {}_{2}F_1(-\frac{1}{\gamma},\frac{\gamma}{4\xi}, 1-\frac{1}{\gamma}, 2\kappa B^2\xi r^{2\gamma}) ~,
\end{equation}
\begin{equation}
B(r)^2= \left( \frac{1}{2\kappa} - r^{2 \gamma} \xi B^2\right)^{-2+\frac{\gamma}{2 \xi}} A(r)^2 ~,
\label{horizon}
\end{equation}
\begin{eqnarray}
\notag V(\phi) & = & 2H \left( \frac{ \phi}{B} \right)^{-\frac{2}{ \gamma}} \left( \frac{1}{2\kappa}-\xi \phi ^2 \right)^{-1+\frac{\gamma}{2 \xi}} \left( 1-2\kappa \xi \phi ^2\right)^{-\frac{\gamma}{4 \xi}}\\
&& -\frac{G}{2}\left( \frac{1}{2\kappa}-\xi \phi ^2 \right)^{-2+\frac{\gamma}{2 \xi}}
\left( \frac{2}{\kappa}-(\gamma^2+4\xi) \phi^2\right)\\ \notag 
&& -\frac{H}{2} \left( \frac{ \phi}{B} \right)^{-\frac{2}{ \gamma}}\left( \frac{1}{2\kappa}-\xi \phi ^2 \right)^{-2+\frac{\gamma}{2 \xi}}
\left( \frac{2}{\kappa}-(\gamma^2+4\xi) \phi^2\right) {}_{2}F_1(-\frac{1}{\gamma},\frac{\gamma}{4\xi}, 1-\frac{1}{\gamma}, 2\kappa\xi \phi ^2)~,
\end{eqnarray}
where $B$, $G$ and $H$ are integration constants. 
This solution is asymptotically AdS and generalizes the previous one, because if we take $\gamma=4\xi$ it reduces to the solution of the case $A(r)=B(r)$. Furthermore, for $\gamma=0$ we recover the static BTZ black hole. On the other hand, in the limit $\phi \rightarrow 0$ the potential tends to a constant $V(\phi) \rightarrow -2G (2 \kappa)^{1-\frac{\gamma}{2 \xi}}=\Lambda$. 

\begin{figure}[h]
\begin{center}
\includegraphics[width=0.6\textwidth]{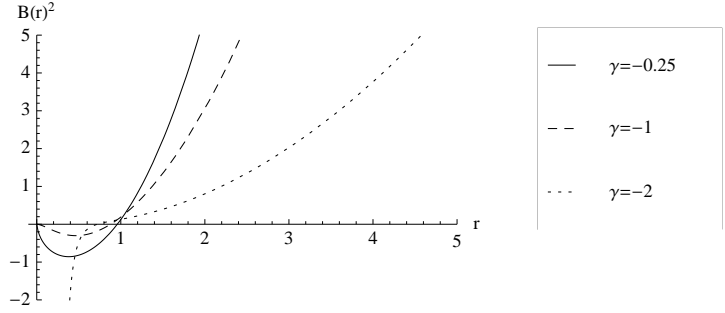}
\end{center}
\caption{The behavior of $B(r)^2$, for $H=-1$, $G=1$, $B=1$, $\kappa=1$, $\xi=-0.25$ 
and $\gamma=-0.25,-1,-2$.} \label{functionn}
\end{figure}

\begin{figure}[h]
\begin{center}
\includegraphics[width=0.6\textwidth]{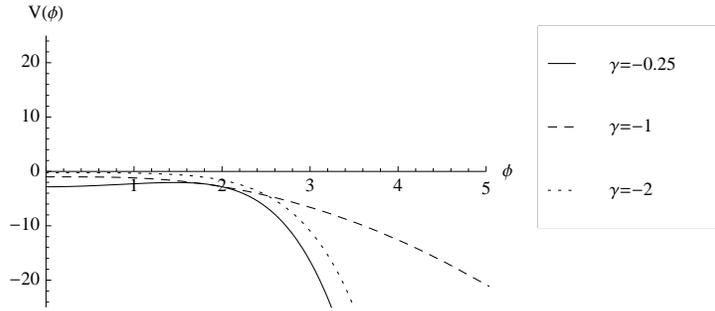}
\end{center}
\caption{The Potencial $V(\phi)$, for $H=-1$, $G=1$, $B=1$, $\kappa=1$, $\xi=-0.25$ 
and $\gamma=-0.25,-1,-2$.} \label{Pot2}
\end{figure}

\begin{figure}[h]
\begin{center}
\includegraphics[width=0.4\textwidth]{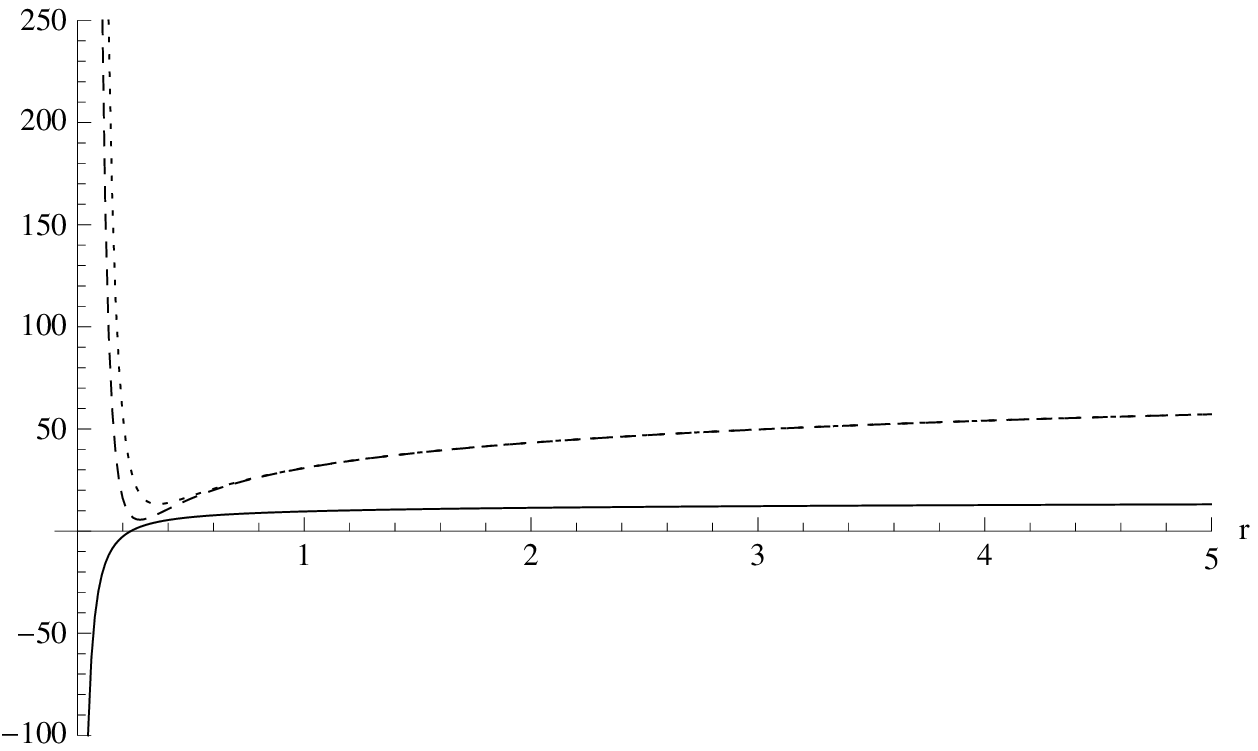}
\includegraphics[width=0.4\textwidth]{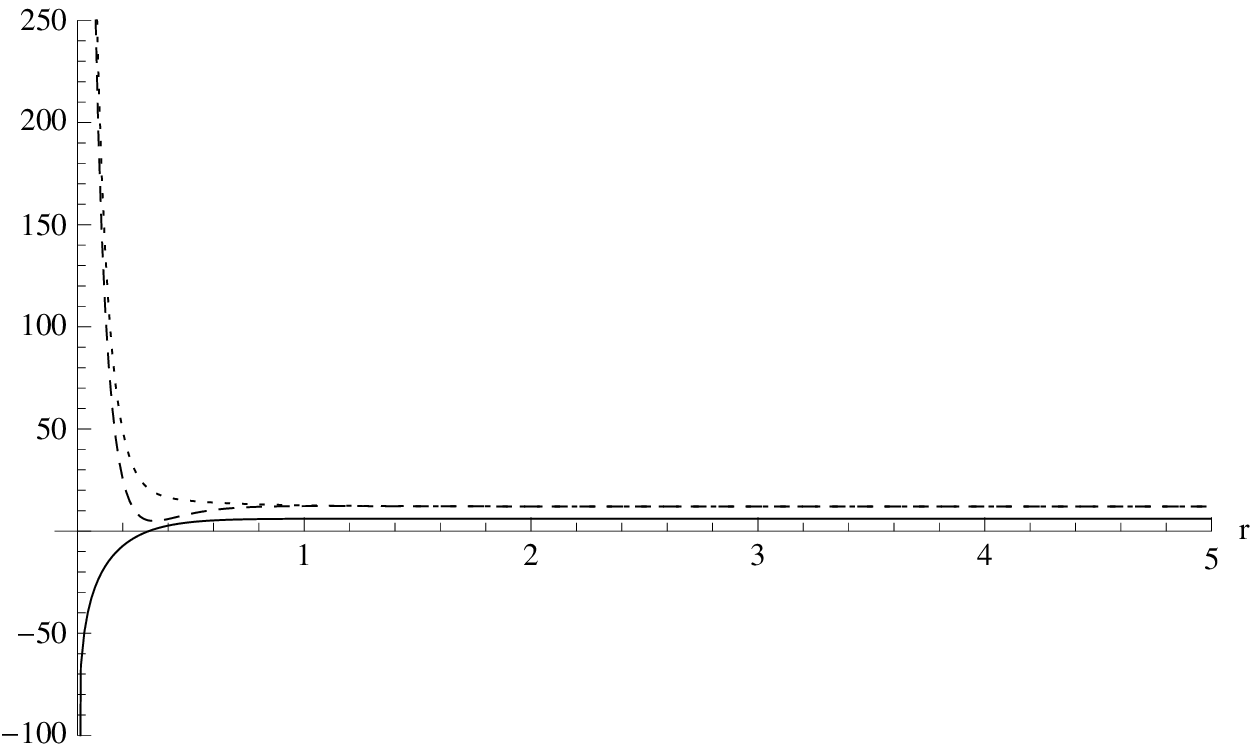}
\includegraphics[width=0.55\textwidth]{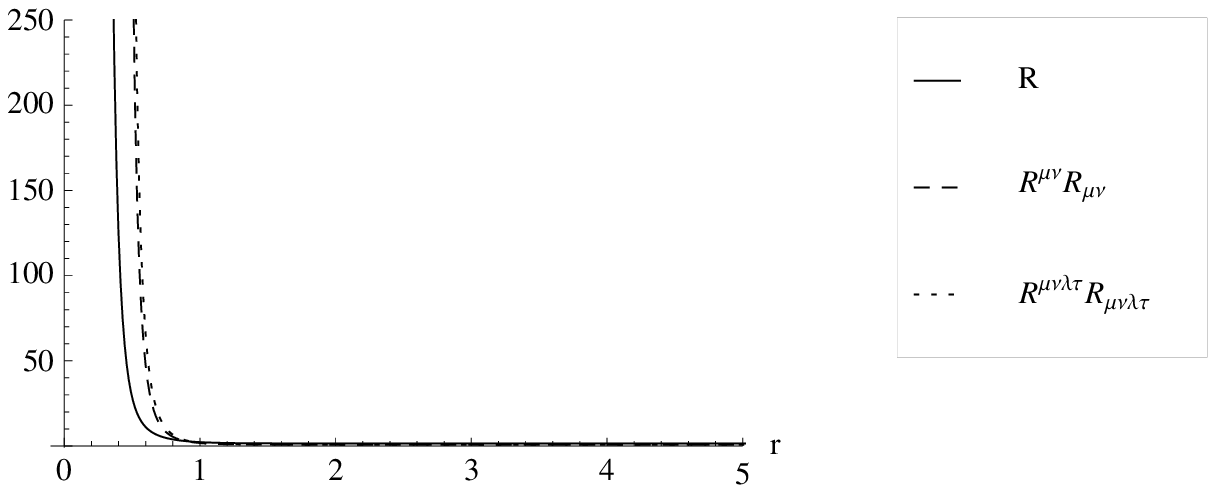}
\end{center}
\caption{The behavior of $R(r)$, $R^{\mu\nu}R_{\mu\nu}(r)$ and $R^{\mu\nu\lambda\tau}R_{\mu\nu\lambda\tau}(r)$ for $H=-1$, $G=1$, $B=1$, $\kappa=1$, $\xi=-0.25$ and $\gamma=-0.25$ (left figure), $\gamma=-1$ (right figure), and $\gamma=-2$ (bottom figure).} \label{figureinvariant}
\end{figure}

\begin{figure}[h]
\begin{center}
\includegraphics[width=0.6\textwidth]{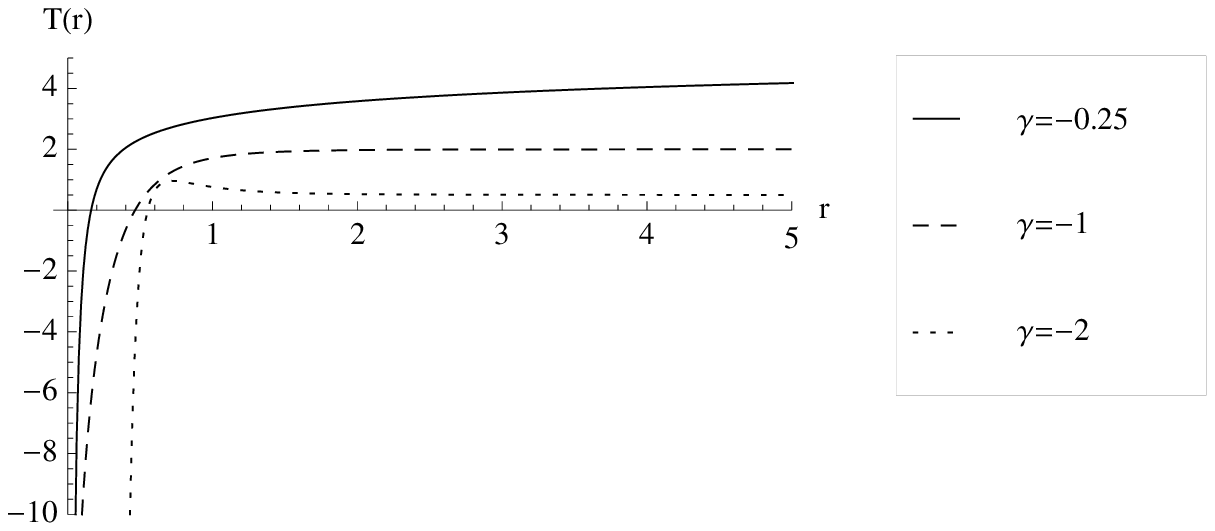}
\end{center}
\caption{The behavior of torsion scalar $T$ as function of $r$ for $H=-1$, $G=1$, $B=1$, $\kappa=1$, $\xi=-0.25$ 
and $\gamma=-0.25,-1,-2$.} \label{torsionRR}
\end{figure}

As in the previous case, we plot the  behavior of the metric function
$B\left( r\right)^2 $ given by (\ref{horizon}), in Fig.~(\ref{functionn}) 
for a choice of parameters $H=-1$, $G=1$, $B=1$, $\kappa=1$, $\xi=-0.25$
and $\gamma=-0.25,-1,-2$. The metric function $B(r)^2$ changes
sign for low values of $r$, signalling the presence of a horizon,
while for $\gamma < 0$ the scalar field is
regular everywhere outside the event horizon and null at large
distances.  In Fig.~(\ref{Pot2}) we show the behavior of the potential, asymptotically ($\phi \rightarrow 0$) it tends to a negative constant (the effective cosmological constant) as in the previous case. Also, we plot the
behavior of $R(r)$, $R^{\mu\nu}R_{\mu\nu}(r)$, and 
$R^{\mu\nu\lambda\tau}R_{\mu\nu\lambda\tau}(r)$  in Fig.~(\ref{figureinvariant}) by using the Levi-Civita connection, and we observe that there is not a
Riemann curvature singularity outside the horizon for  $\gamma=-0.25,-1,-2$. Also, we observe a Riemann curvature singularity at $r=0$ and the torsion scalar is singular at $r=0$ for all the cases considered. Asymptotically, the torsion scalar goes to $-2\Lambda$ since this spacetime is asymptotically AdS, see Fig.~(\ref{torsionRR}). Therefore, we have shown that there are three-dimensional black hole solutions with scalar hair in Teleparallel Gravity.

\section{Final Remarks}
\label{conclusions}

Motivated by the search of hairy black holes solutions in theories based on torsion, we have considered an extension of three-dimensional TEGR with a scalar field non-minimally coupled to the torsion scalar along with a self-interacting potential, and we have found three-dimensional asymptotically AdS black holes with scalar hair. These hairy black holes are characterized by a scalar field with a power-law behavior and by a self-interacting potential, which tends to an effective cosmological constant at spatial infinity. We have considered two cases $A(r)=B(r)$ and $A(r)\neq B(r)$. In the first case the scalar field depends on the non-minimal coupling parameter $\xi$, and it is regular everywhere outside the event horizon and null at spatial infinity for $\xi < 0$, while for $\xi = 0$ we recover the non-rotating BTZ black hole. In the second case the scalar field depends on a parameter $\gamma$, and it is regular everywhere outside the event horizon and null at spatial infinity for $\gamma < 0$, this solution generalizes the solution of the first case, which is recovered for $\gamma=4\xi$. Furthermore, for $\gamma = 0$ we recover the non-rotating BTZ black hole. Moreover, the analysis of the Riemann curvature invariants and the torsion scalar shows that they are all regular outside the event horizon. In furthering our understanding, it would be interesting to study the thermodynamics of these hairy black hole solutions in order to study the phase transitions. Work in this direction is in progress.

\acknowledgments 
This work was funded by Comisi\'{o}n
Nacional de Ciencias y Tecnolog\'{i}a through FONDECYT Grants 11140674 (PAG),
1110076 (JS) and 11121148 (YV) and by DI-PUCV Grant 123713
(JS). P.A.G. acknowledge the hospitality of the
Universidad de La Serena.

\vskip 7cm

\end{document}